\documentclass{article}
\usepackage[utf8]{inputenc}
\usepackage[margin=1in]{geometry}
\usepackage{setspace}
\usepackage{amsmath}
\usepackage{authblk}

\title{Diagnosis of systemic risk and contagion across financial sectors}
\author[1]{Sayuj Choudhari}
\author[2]{Richard Zhu}

\affil[1]{Adlai E. Stevenson High School}
\affil[2]{Department of Computational and Applied Mathematics, University of Chicago}

\date{December 2020}

\usepackage{natbib}
\usepackage{graphicx}

\begin{document}

\maketitle

\pagebreak

\tableofcontents

\pagebreak

\section{Introduction}
	\indent
    As is expected with hindsight, there were obvious signs that significant systemic risk had been pent-up in the economy prior to the Great Financial Crisis (GFC) of 2007-2009.
    Deregulation beginning nearly a decade prior enabled many financial institutions to take on massive leverage.
    From May 2000 to June 2003, the Federal Reserve cut interest rates from 6.5\% to 1\% over in response to the crash of the dot-com bubble, multiple accounting scandals, and the terrorist attacks of September 11th.
    While this provided a short term fix, cut interest rates replaced the dot-com bubble with a real estate bubble.
    Lenders originated loans to many people financially unfit to buy homes, passing them onto the burgeoning secondary market for originating and distributing subprime loans (Investopedia, 2020).

    The repeal of the Glass-Steagall Act in 1999 loosened the rules on the abilities of banks and brokerages, allowing them to both take deposits and make loans, as well as underwrite and sell securities.
    Rickards (2012) explains the danger of this: these institutions could originate loans, repackage the cash flows from the loans as securities, and pass off these mortgage-backed securities to other buyers.
    Because these novel securities paid huge fees to the originators, more and more institutions decided to get in on the game.
    The SEC also lowered net asset requirements for Goldman Sachs, Merrill Lynch, Lehman Brothers, Bear Stearns, and Morgan Stanley, which enabled these institutions to increase leverage massively.
    Repackaging MBSs became a shell game, creating a scheme where trillions of dollars of securities held on balance sheets could be tethered to the performance of mere billions of mortgages.

    In normal times, it is assumed that financial institutions operating in non-overlapping sectors have complementary and distinct outcomes, typically reflected in mostly uncorrelated outcomes and asset returns.
    Such is the reasoning behind common "free lunches" to be had in investing, like diversifying assets across equity and bond sectors.
    Unfortunately, the recurrence of crises like the GFC demonstrate that such convenient assumptions often break down, with dramatic consequences for all financial actors.
    The emergence of so-called systemic risk (exemplified by failure in one part of a system spreading to ostensibly unrelated parts of the system) is thus far clearly evinced in the deregulation narrative.
    But is it possible to diagnose and quantify the emergence of systemic risk in financial systems?

    In this study, we focus on two previously-documented measures of systemic risk that require only easily-available time series data (eg monthly asset returns): cross-correlation and principal component analysis.
    We apply these tests to daily and monthly returns on hedge fund indexes and broad-based market indexes, and discuss their results.
    We hope that a frank discussion of these simple, non-parametric measures can help inform legislators, lawmakers, and financial actors of potential crises looming on the horizon.
	
	Section 2 provides a review of literature on previous research and methods used to analyze and measure systemic risk. The measures and methods used in this study are described in Section 3. The data tables and plots of the results from the systemic risk measurement analyses conducted is presented in Section 4, as well as a brief explanation of the meaning of the data found. The analysis of this data with the conclusions of this study are given in Section 5.

\section{Literature Review}

Following the work of Andrew Lo (2010) we use the definitions of systemic risk and the crises it causes that are proposed by De Bandt and Hartmann (2000):

\vspace{5mm} 

A systemic crisis can be defined as a systemic event that affects a considerable number of financial institutions or markets in a strong sense, thereby severely impairing the general well-functioning of the financial system. While the “special” character of banks plays a major role, we stress that systemic risk goes beyond the traditional view of single banks’ vulnerability to depositor runs. At the heart of the concept is the notion of “contagion”, a particularly strong propagation of failures from one institution, market or system to another. 

\vspace{5mm} 

We use the the requirements for systemic risk measure described by Brunnermeier et al. (2009) which is:

\vspace{5mm} 

A systemic risk measure should identify the risk on the system by individually systemic institutions, which are so interconnected and large that they can cause negative risk spillover effects on others, as well as by institutions which are systemic as part of a herd.

\vspace{5mm} 

The statistical tests we conduct in this research follow those definitions for the description and measurement of systemic risk. The importance of investigating the connectedness of financial institutions/ sectors in the study of systemic risk, which is done in this study,  is shown in Allen (2001) where the interlinkages between financial institutions and its effects are identified. These evaluations of interconnectedness within the finance world are further studied by many other researchers. Barunik (2018) evaluates the dynamics of U.S. financial institutions for interlinkages on the short and long term and their varying effects on the economy. The book “Financial and Macroeconomic Connectedness: A Network Approach to Measurement and Monitoring” by Francis Diebold and Kamil Yilmaz 2015 also develops strategies to identify relations between financial institutions to create a deep understanding of the global financial network and the causes and effects of financial interlinkages. Aikman (2017) proposes the use of many types of economic statistics for various financial institutions to map out areas of vulnerability within the U.S. economy.

Seeing that the interconnectedness between financial institutions was becoming an increasingly popular area of study for systemic risk, we now had to look into a specific financial institution/ sector to focus our tests on. De Bandt and Hartmann’s definition of systemic risk  emphasized that banks, although seemingly playing a significant role in market crashes and systemic failures, are not the only cause of such crises as systemic risk goes beyond a singular institution or system making it necessary to consider other parts of the market as well. Thus, we looked into what financial sectors may have played important roles in crises and from which of these major sectors systemic risk may have stemmed from. Multiple sources of literature were found investigating or citing hedge funds as an early indicator of systemic risk in the recent 2008 market crash. Lo (2008) describes and evaluates the relation between systemic risk and hedge funds during that time period, and furthered his research into showing possible “early warning signs” seen within the hedge fund sector of the coming market crash. Systemic risk by hedge funds imposed on the banking sector (commonly believed as the key role in systemic failures) is studied by Chan (2005), demonstrating the possibility of systemic risk within banking in 2008 to have stemmed from the hedge fund sector. Furthermore, the effect of hedge funds on  counterparty credit risk management (CCRM, systemic risk and market crash prevention measures) is examined in Kambhu, Schuermann, and Stiroh (2007), where the different structures and operations of hedge funds are shown to challenge CCRM and increase systemic risk. These works along with many others, led us to focus our testing of systemic risk on the hedge fund sector and investigate the statistical data between hedge funds with other sectors as well as within different hedge funds themselves.

	Our use of Principal Component Analysis (PCA) to measure systemic risk is closely based off of Lo (2010) in which hedge funds, banks, brokerages, and insurers were tested for systemic risk using the return data of the indexes of the mentioned financial sectors. Similar work is also seen in Kritzman, Li, Page, Rigobon (2011) where principal components are used to measure the connectedness of institutions within a market as well as the market fragility. This group of researchers extends their work in 2012 to show the use of PCA to evaluate the risk specific entities have on the market, how vulnerable specific entities are to systemic risk.
	The use of cross-correlations to analyze systemic risk in this study is also complementary to that of Lo (2001) in which the cross-correlation coefficients between 2 types of financial data are used to identify possible systemic risk within the market. The single comparison used in cross-correlations, which cannot be done in PCA, is helpful in creating more highly defined mappings of systemic risk and connectedness within the financial markets, as they allow for the analysis of a singular sector/ institution separately against various others. Chiang, Jeon, and Li (2007) also implemented the methods of correlations on the Asian market to discover different financial phases of the asian market crash. Also along the same lines, Getmansky, Lo, and Makarov (2004) examine serial correlation within hedge fund returns and the high correlation levels being likely explained by the illiquidity exposure within the hedge fund industry, which may have initiated systemic risk. The timing and results of this work showing illiquidity exposure risk within the hedge fund industry only a few years before the financial crisis of 2008 further motivated us to emphasize our work towards the movements of hedge funds.

\section{Systemic Risk Measures}

\subsection{Principal Component Analysis}

Principal component analysis (hereafter, known as PCA) involves the eigenvalue decomposition between data sets/ components, which calculates the amount of variation brought by each data set/ component (see Muirhead, 1982 for an exposition of PCA). In context, when most of the variation is accounted for in the first few components this would mean that the components are more correlated as there is less variation among them and their movements are similar. Therefore, when applying PCA to finance, each component can be a time-series return data on an asset, and detect increased correlations among assets over certain time periods, which in other words is measuring systemic risk and the exposure each asset has on the other. In this study, PCA is conducted on time-series data for hedge funds with the S\&P 500 to detect possible systemic risk between hedge funds and the overall US economy, as well as just on various hedge fund sectors themselves to detect possible systemic risk within the hedge fund industry itself.

The first step in the PCA of this study is to create a returns matrix of the data (hereafter, known as $\mathbf{R}$), where each row is an asset (In this study: LAB hedgefund index, banks, brokerages, and insurers) and each column is each row (asset) is  return data at a certain step in the time-series, $\mathbf{R}$($n$, $k$). The mean return of each asset is then subtracted from each respective row creating the (insert name of matrix type) matrix of $\mathbf{R}$ (Matrix $\mathbf{M}$):

$$ \mathbf{M}_{n, k} =
\begin{bmatrix}

    \mathbf{R}_{1,k} - \overline{\mathbf{R}_{1,k}}\cr
    \mathbf{R}_{2,k} - \overline{\mathbf{R}_{2,k}}\cr
    \vdots \cr
    \mathbf{R}_{n-1,k} - \overline{\mathbf{R}_{n-1,k}}\cr
    \mathbf{R}_{n,k} - \overline{\mathbf{R}_{n,k}}\cr
\end{bmatrix} $$

Matrix M, now represents the variance of each asset is return data from the mean of the returns. Thus, when M is multiplied by the transposed version of itself, $\mathbf{M}^{T}$, the covariance matrix $\Sigma$ is given:

$$\mathbf{Var} [{R}_{t}]  \equiv   \Sigma = \mathbf{M} \mathbf{M}^{T} =
\begin{bmatrix}
   
    \mathbf{M}_{1,k} \cdot \mathbf{M}^{T}_{k, 1} & \hdots & 
    \mathbf{M}_{1,k} \cdot \mathbf{M}^{T}_{k, n} \cr
    \vdots && \vdots \cr
    \mathbf{M}_{n,k} \cdot \mathbf{M}^{T}_{k, 1} & \hdots & 
    \mathbf{M}_{n,k} \cdot \mathbf{M}^{T}_{k, n} \cr

\end{bmatrix}$$

As seen above, each component of $\Sigma$ represents the correlation between the two assets that are represented in that location. For example, the component (n,k) in $\Sigma$ represents the value $n_{1}k_{1} + n_{2}k_{2} +....+ n_{M}k_{M}$ where $n_{t}$ and $k_{t}$ is the variance of return for assets n and k respectively over a time step. Looking back at the calculation of that component $(n, k)$ it is seen that the component can indicate the typical movements between the two assets. Considering that this calculation is done for all components of $\Sigma$, $\Sigma$ is representative of all the movements between all the assets. Thus, when we take the eigenvalue decomposition of $\Sigma$ we can observe the weightage of each assets movements, this can be expressed as:

$$ \mathbf{Var} [{R}_{t}] \equiv \Sigma  = \mathbf{Q} \mathbf{\Theta} {\mathbf{Q}^{\prime}} ,  \Theta = 
\begin{bmatrix}

    {\theta}_{1} & 0 & \hdots & 0 \cr
    0 & {\theta}_{2} && 0 \cr
    \vdots && \ddots & \vdots \cr
    0 & \hdots & 0 & {\theta}_{N}
\end{bmatrix} $$

In the above equation, matrix $\mathbf{\Theta}$ would have the eigenvalues of $\mathbf{\Sigma}$ along it's diagonal along with Matrix $\mathbf{Q}$ containing the corresponding eigenvectors. Thus, we get a set of eigenvalues, where each eigenvalue finds the total variation accounted for in each principal component  which is essentially reflecting the importance of each component is movements for all assets analyzed. Once these eigenvalues are then normalized to a sum of 1, we are given the fractional eigenvalues corresponding with each principal component which allow us to analyze where the variation of the assets is movements is accounted for. One issue that arises with the use of PCA is that the covariance matrix $\Sigma$ must be estimated:

$$ \widehat{\mathbf{\Sigma}} \equiv  
\frac{1}{T-J}
\sum_{t=1}^{T} 
(\mathbf{{R}_{t} - \overline{\mathbf{R}}})
(\mathbf{{R}_{t} - \overline{\mathbf{R}}}) ^ \prime
$$

is singular when the number of assets $J$ compared as larger than the amount of time-series data points $T$ available for each asset. Thus, the number of assets we compare is limited to only hedge funds, banks, brokerages, and insurers (Lo, 2010), and the windows we use for each PCA calculation is relatively large (30 datapoints) to maximize degrees of freedom and create more accurate results.

So, using Principal Component Analysis, if the fractional variation of the first or first few eigenvalues was very high and quickly approached 1, then the movements of all assets compared are very correlated as most variation is accounted for in only one or a few principal component(s) and pose a possibility of having systemic risk. On the other hand if the fractional variation of the assets does not quickly approach the noise around 1 within the first few components, then the variation is spread across all assets and movements are therefore uncorrelated and do not indicate systemic risk.

In this study, the fractional variation of the first component is calculated over a time period of (TBD) as it is best representative of the correlation between all assets whereas after 3 components are considered the fractional variation goes into the noise approaching 1. Mapping out the data of each PCA on a (TBD-length) time period over the entire time-series of data, we can detect the time-periods when fractional variation is highest and systemic risk is likely present.

\subsection{Cross-correlations}

In order to determine the influence or exposure of asset movements on each other, cross correlations can calculate coefficients that determine the extent as to how much one asset is movements are reflected in another. The reason as to why similar movements over a period of time are a result of correlations rather than chance can be derived from the martingale model: the idea that all information in an informationally efficient environment is accounted for in an assets price making future prices unpredictable and random (Samuelson, 1965). However, this informationally efficient environment required in the martingale model is never fully achieved as the market can never be perfectly efficient, and therefore predictable points or market friction can occur causing correlation/ exposure between assets. This friction is an example of assets being influenced/ exposed to others as the movements of one asset are creating the movements of another, rather than a result of chance. Furthermore, in a time period of systemic risk, these correlations/ exposures between assets must be increasing as the martingale model fails and market friction increases (Lo, 2010). With this direct relationship between systemic risk and correlation between assets, testing cross-correlations to find levels of correlation between assets is a possible way to find systemic risk levels between those assets as well. Cross-correlation also allows comparison of one asset is movements to a lagged version of the other. This becomes helpful in the case that an asset is movements are a lagged reflection of another, as the effects of market friction and systemic risk may take a certain time period to set in.

The cross-correlation coefficient is the variable within a cross-correlation calculation that indicates the levels of correlation between two assets with a possible time lag ($l$). The data necessary for this calculation is a time-series of return data on an asset with a constant time-step. The first step is to create a data set for each asset that is the variance of that asset is actual returns over a time period from the mean returns over that time period:

$$ \mathbf{Var}[\mathbf{A}] = \mathbf{A} - \overline{\mathbf{A}} $$

$$ \mathbf{Var}[\mathbf{B}] = \mathbf{B} - \overline{\mathbf{B}} $$

Using these two new data lists for the two assets one of the asset lists is converted into a convoluted matrix and multiplied by the data list of the second asset. This finds the sum of all covariances between the two assets at any time lag in the range of the time period tested for $l \in [0, $k$)$ where $k$ is the length of the time period:

$$
C = A*B =
  \begin{bmatrix}
    A_1 & 0 & \hdots & 0 & 0\cr
    A_2 & A_1 && \vdots & \vdots \cr
    A_3 & A_2 & \hdots & 0 & 0\cr
    \vdots & A_3 & \hdots & A_1 & 0\cr
    A_{k-1} & \vdots & \ddots & A_2 & A_1 \cr
    A_{k} & A_{k-1} && \vdots & A_2\cr
    0 & A_{k} & \ddots & A_{k-2} & \vdots \cr
    0 & 0 & \hdots & A_{k-1} & A_{k-2} \cr
    \vdots & \vdots && A_{k} & A_{k-1} \cr
    0 & 0 & 0 & \hdots & A_{k} \cr
  \end{bmatrix}
  \begin{bmatrix}
    B_1 \cr
    B_2 \cr
    \vdots \cr
    B_{k-1} \cr
    B_{k} \cr
  \end{bmatrix}
$$

As seen in the equation above, a vector of length 2$k$ - 1 is created, with the center component being the sum of covariances at a time lag of 0 (e.g. considering asset A and asset B the center of this vector would be $A_{1}B_{1}+A_{2}B_{2}+...+A_{k}B_{k}$ where $k$ is the length of the time period tested over). Therefore, all vector components ahead of that center component are truncated to a new vector that is the same length as the time period $\mathbf{D}$ = [$C_0$, $C_{1}$, ... , $C_{k-2}$, $C_{k-1}$] and contains all covariance sums between the two assets for $l \in [0, k)$. Now considering that the covariance sum with $l = 0$ was composed of covariances on all time steps in the time period, while the covariance sum with $l = k - 1$ was composed of only one covariance calculation. Thus, each covariance sum is then divided by the amount of time steps that are accounted in its calculation to find the average covariance over a time step for each lag, giving all components of the covariance vector comparable values. Each component is now divided by the standard deviation of the two assets to provide the cross-correlation coefficient($r$) at each time lag:

\begin{center}

$r = \frac {D} {[1, 2, ..., k-1, k] * \sigma(Var[\mathbf{A}], Var[\mathbf{B}])}$

\end{center}

So, the final result of all these calculations gives cross-correlation coefficients between the two assets on any given time lag over the time period tested for. Systemic risk over a time period can then be detected with this test as an increase in cross-correlation (which was defined earlier as having a direct relationship with systemic risk) can now be detected over that time period. In this study, a lag of 1 day is used to identify cross-correlation coefficients as it most accurately represents the lag caused by market friction. To evaluate the cross-correlation with a lag as small as 1 day, the time-step is also set at daily and the time-period is set at 90 days or one quarter. This time-period window of 90 days is best as it is just long enough to eliminate the possibility of randomness and chance causing similar movements, allowing the highest definition spikes in cross-correlation to be observed as a legitimate possibility of systemic risk.

\section{The Data}

\subsection{Hedge Funds}

The Hedge Fund data used in this study comes from the Credit Suisse Liquid Alternative Beta Index (hereafter, known as “LAB Index”) and contains daily pricing information and can therefore be used to find return data over any period of time from January 1998 to June 2020. It consists of liquid securities and aims to reflect the returns of the overall hedge fund industry strategies that use liquid and tradable instruments. The LAB Index develops a measure for returns on the entire hedge fund industry through including hedge fund data from 5 different Credit Suisse Hedge Fund Indices that differ in strategy: the Event Driven Liquid Index (hereafter, known as “EDL Index”), Global Strategies Liquid Index (hereafter, known as “GSL Index”), Long/ Short Liquid Index (hereafter, known as “LSL Index”), Managed Futures Liquid Index (hereafter, known as “MFL Index”), and Merger Arbitrage Liquid Index (hereafter, known as “MAL Index”). The LAB Index  takes into account the different weights each of these indices have in the overall hedge fund industry. This weighting is reflective of each strategy’s respective weight in the Credit Suisse Hedge Fund Index.
	
This research also includes the separate pricing data for each of the 5 LAB sub-indices. This separate data for each hedge fund strategy helps to run tests for systemic risk within the hedge fund industry itself through providing individual data on these 5 hedge fund sectors. The hedge fund sectors that are part of these indices are all susceptible to systemic risk from the other sectors as they all are part of the LAB index, which as known earlier, contains liquid hedge fund strategies that are exposed to economic movements/ events occurring. Thus, systemic risk within the hedge fund industry can be seen through these 5 indices and should be measured. Furthermore, it is necessary to establish a definition for each of these indices as to help provide reasoning for the results of the systemic measurement tests conducted in Section 5. As according to Credit Suisse, the following are the definitions for each of the indices /indice strategies:

\vspace{3mm}

\onehalfspacing

EDL Index: Develops trading strategies to create returns and solid  out of global events (e.g. elections, 

corporate situations/ activities,  mergers and acquisitions deals, etc.) 


\vspace{3mm}

GSL Index: Reflects the return of all remaining hedge fund strategies not defined as Long/Short or Event 

Driven

\vspace{3mm}

LSL Index: Strategy that takes a long term position on underpriced equities and also selling short equities 

that are expected to decline in value

\vspace{3mm}

MFL Index: Uses a pre-defined quantitative methodology to invest in a range of asset classes including: 

equities, fixed income, commodities and currencies

\vspace{3mm}

MAL Index: Uses using a pre-defined quantitative methodology to invest in a liquid, diversified and 

broadly representative set of announced merger deals

\doublespacing

\subsection{S\&P 500}
The S\&P 500 data is obtained from investing.com and contains the daily pricing and return values from January 2006 to June 2020. Due to this data only containing returns from 2006 onwards, while the hedge fund data from Section 4.2 goes back to 1998, the hedge fund data for days that do not correspond with the S\&P 500 data is truncated during the tests that compare the two different financial sectors. The S\&P 500 is an asset-weighted index containing the top 500 companies in the US market, and is used as a general evaluation for the US economy as a whole. Similar to the hedge fund data, the daily information provided by investing.com allows to calculate return values over any time period within the range of the data, meaning monthly data can be created in cases that the second data set compared with contains monthly data. Data that represents the entire US economy and its movements is relevant in this study as it provides the means to measure systemic risk of the hedge fund industry on the overall US economy. More specifically it also allows testing each of the LAB Index is 5 subindices for possible systemic risk with the US economy. From those tests, we can determine which hedge fund strategies might be susceptible to or creating systemic risk within an economy. Overall, the S\&P 500 data provides a gauge of the US economy which is important to test with the hedge fund data as it can evaluate the relations and movements between hedge funds and the rest of the financial/ corporate world.

\subsection{Banks}

The bank data is obtained from the Wharton School of Economics is WRDS service with the CRSP daily stock price database and contains daily pricing and return information from February 1990 to June 2020 of banks with SIC codes from 6000-6199. The return data is then used along with the market capitalization of each firm to create a market-cap weighted index of returns for the banks. This banking index will serve as a measurement of the entire banking sector. The data is truncated for the same date range from January 2006 to June 2020 as mentioned in 4.2. Similar to all other data in this study, the daily returns from the banking index can be used to create return data for any length time period within the time range. This banking data is important as it provides the banking sector component of the financial industry that when combined with other components from this study (hedge funds, brokerages, and insurances) can be tested for systemic risk within the entire economy using PCA (statistical test explained in 3.2).

\subsection{Brokerages}

The brokerage data is obtained from the Wharton School of Economics is WRDS service with the CRSP daily stock price database and contains daily pricing and return information from February 1990 to June 2020 of brokerages with SIC codes from 6200-6299. The return data is then used along with the market capitalization of each firm to create a market-cap weighted index of returns for the banks. This banking index will serve as a measurement of the entire brokerage sector. The data is truncated for the same date range from January 2006 to June 2020 as mentioned in 4.2. Similar to all other data in this study, the daily returns from the brokerage index can be used to create return data for any length time period within the time range. This brokerage data is important as it provides the brokerage sector component of the financial industry that when combined with other components from this study (hedge funds, banks, and insurances) can be tested for systemic risk within the entire economy using PCA (statistical test explained in 3.2).

\subsection{Insurers}

The insurance data is obtained from the Wharton School of Economics is WRDS service with the CRSP daily stock price database and contains daily pricing and return information from February 1990 to June 2020 of insurances with SIC codes from 6300-6499. The return data is then used along with the market capitalization of each firm to create a market-cap weighted index of returns for the insurances. This insurance index will serve as a measurement of the entire insurance sector. The data is truncated for the same date range from January 2006 to June 2020 as mentioned in 4.2. Similar to all other data in this study, the daily returns from the insurance index can be used to create return data for any length time period within the time range. This insurance data is important as it provides the insurance sector component of the financial industry that when combined with other components from this study (hedge funds, banks, and brokerages) can be tested for systemic risk within the entire economy using PCA (statistical test explained in 3.2).

\section{Results}

In this section, the 2 systemic risk measurement methods discussed in Section 3 are applied to the S\&P 500 and Hedge Fund data mentioned in Section 4. Section 5.1 displays the results of cross-correlation calculations conducted on the assets is return data, and Section 5.2 gives the results of PCA conducted on that data as well.

\subsection{Principal Component Analysis}

The principal component analysis (hereafter referred to as PCA) is conducted between Hedge Funds, Brokerages, Banks, and Insurances (all key financial asset types in the market). The 1st fractional eigenvalue is the ideal data point to record as it is the most accurate measure of the concentration of variation between assets is movements as the 1st fractional eigenvalue is too statistically unreliable and the fractional eigenvalues go into the noise approaching 1 at the 3rd. Being the most accurate measure for concentration of variation in asset movements it is then also the ideal systemic risk indicator as increased concentration in the variation of assets means increasingly similar movements and market friction. With the 4 components of data being considered as independently moving assets, the expected PCA fractional eigenvalues is an equal spread of variation among all assets/ components (fractional eigenvalues from 1-4 would be: .25, .5, .75, 1). However, this statistically expected data is highly unlikely, and any abnormal similarities in movements should create an increase in the value of the first component, indicating systemic risk across the entire financial market as the assets being compared do span the entire financial industry. The PCA calculations and plot use the return data for the main LAB hedge fund index, as well as market-cap weighted indices of brokerages, banks, and insurances (for further explanation on the indices refer to Section 3.1) over the same 174-month time period from January 2006 to June 2020. Figure 3 is the plot for the second fractional eigenvalues for the PCA on the 4 asset indices mentioned above, using 2 week return periods and 4 month (16 week) windows for each calculation over time.

\begin{figure}[h!]
\centering
\includegraphics[scale=.9]{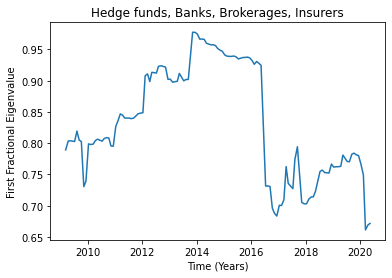}
\caption{Plot of first fractional eigenvalues of return data from indices of Banks, Brokerages, Insurers, and the LAB hedge fund index (refer to sections 4.1, 4.3, 4.4, and 4.5 for further detail on data used) over rolling windows of 30 time periods from over a 174 month range from January 2006 to June 2020.}
\label{fig:Hedge Funds, Banks, Brokerages, and Insurers PCA}
\end{figure}

The plot from Figure 3 shows that from January 2006 to mid 2008 (periods 0 to ~mid 30’s), the first fractional eigenvalue remains between .65 and .8. Although for complete statistical independence the first fractional eigenvalue of 4 assets compared should be .25 it is understandable to have that high value above .5 when comparing movements of different financial industries as all industries should be moving in the same general direction as the entire economy to some extent. Also considering that during that 2.5 year time period there was no market collapse, the  .65 to .8 range recorded in that time period helps to set an expected value of what the first eigenvalue should be in a normal market situation. Looking at the crash times from mid 2008 to 2012 (time periods ~mid 30’s to ~mid 70’s) shows a noticeable spike to values above .9 and even above .95, which are significantly outside of the normal range of .65 to .8. This spike only begins to trend downward to the normal range during the periods following 2012 (time periods mid-70’s on onward) suggesting that the significantly high first fraction eigenvalues within the market crash time are related to the crash and not a cause of an overall market change. Additionally, a high first fractional eigenvalue suggests that a lot of the variation in asset movements can be measured within the first principal component, showing that the asset movements are very highly correlated. This significantly high correlation is another indicator of high market friction (See section 3.2 for reasoning) between the 4 assets compared during the market crash period. Like in the results of the cross-correlations (section 5.1)  this spike in market friction can be representative of a spike in systemic risk as well as the correlations between all assets would mean that each asset can pose a threat to all others, making the group of industries very fragile and vulnerable to a systemic failure which was observed during the time period. Mainly, the PCA results recorded in this test show a spike in fractional eigenvalue during the market crash from 2008-2012, which can be indicative of a systemic risk within the network of assets tested for (hedge funds, banks, brokerages, and insurers) that needs to be eliminated in the future.

\subsection{Cross-correlations}

This calculation of cross-correlations of Hedge Funds and the S\&P 500 US market measure, display the cross-correlation coefficient between 2 of the given assets with a lag of one day to account for possible delay in systemically related movements. This cross-correlation method is conducted on the main LAB index as well as its 5 sub-indices against the S\&P 500 over a 174-month period from January 2006 to June 2020. Figure 1 provides the plots of cross-correlation coefficients over that time period between the LAB index and S\&P 500.

\begin{figure}[h!]
\centering
\includegraphics[scale=.5]{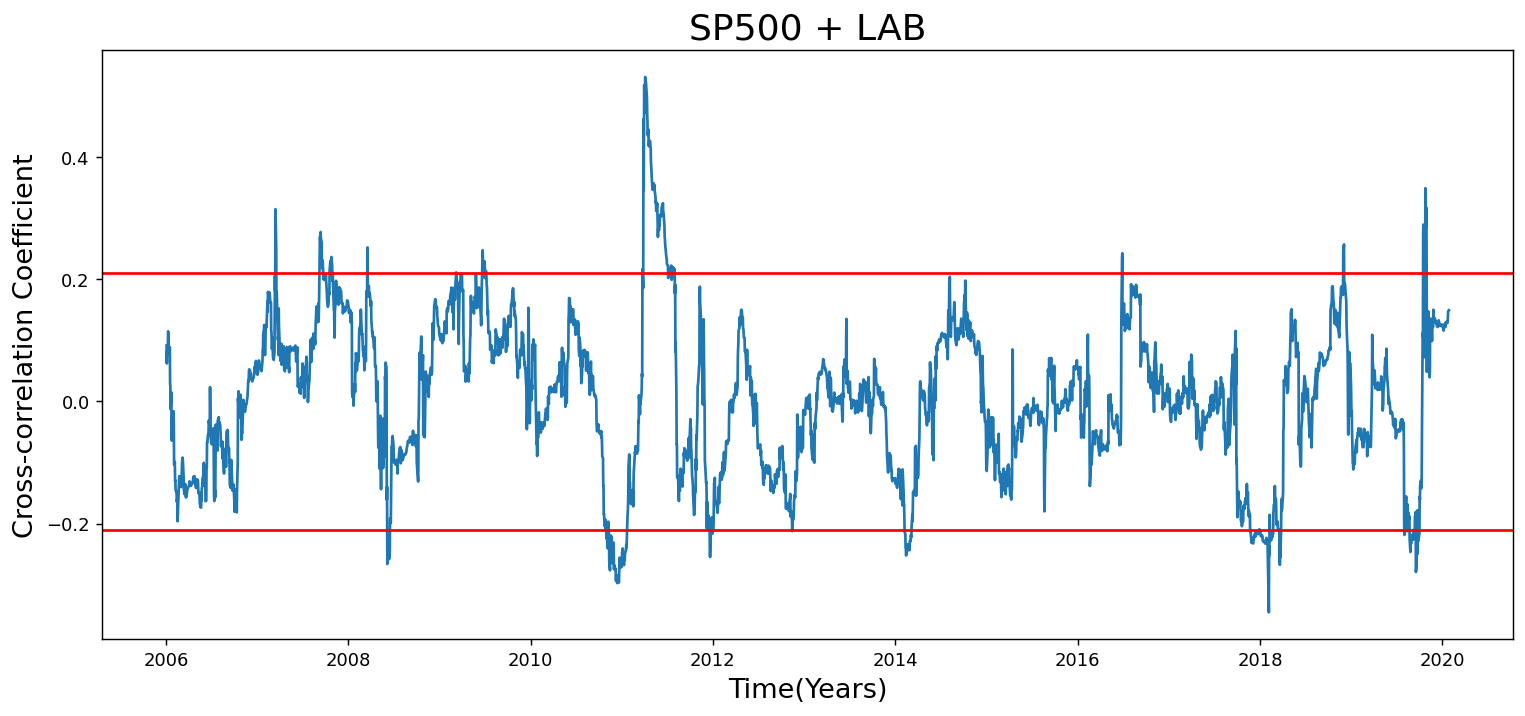}
\caption{Cross-correlation coefficients plot (represented by blue line) with significance bands (red line) between the LAB index and S\&P 500 over the 174-month time frame from January 2006 to June 2020.}
\label{fig:SP500 & LAB cross-correlations}
\end{figure}

As seen in Figure 1, the cross-correlation levels between the LAB index and S\&P 500 show movements generally being part of the statistical noise within the significance bands of ~21\% (Calculated with lags = $\mathbf{l}$ = 90, significance bands = $\pm 2 \sqrt{l}$). The outliers of this trend are most frequent in the early 2007 to 2008 time period, the same time as the market crash. During this period, the cross-correlation between the two assets reached high 20\% to low 30\%, which is statistically significant. Soon after in mid 2008, it lowered from this significantly high cross correlation to a statistically significant negative cross-correlation of -28\% for a short period of time. Regardless of the direction of the cross-correlation coefficients between the two assets over that time period, the plot having frequent statistically significant cross-correlation between the two assets over that time period, may show that market friction is present during that time period along with significant, possibly dangerous levels of systemic risk. Thus, it is seen that systemic risk may have existed between the overall Hedge Fund Industry and the US market (represented by the S\&P 500) during the 2007-2008 market crash. Considering these results, we can then plot the cross-correlation coefficients between the 5 subindices which may help to pinpoint the specific hedge fund sector that is responsible for the possible systemic risk posed by the entire hedge fund industry on the US market. Figure 2 shows the cross-correlation plots between the 5 subindices and the S\&P 500 over the same 174-month time period plotted in Figure 1.

\begin{figure}[h!]
\centering
\includegraphics[scale= .25]{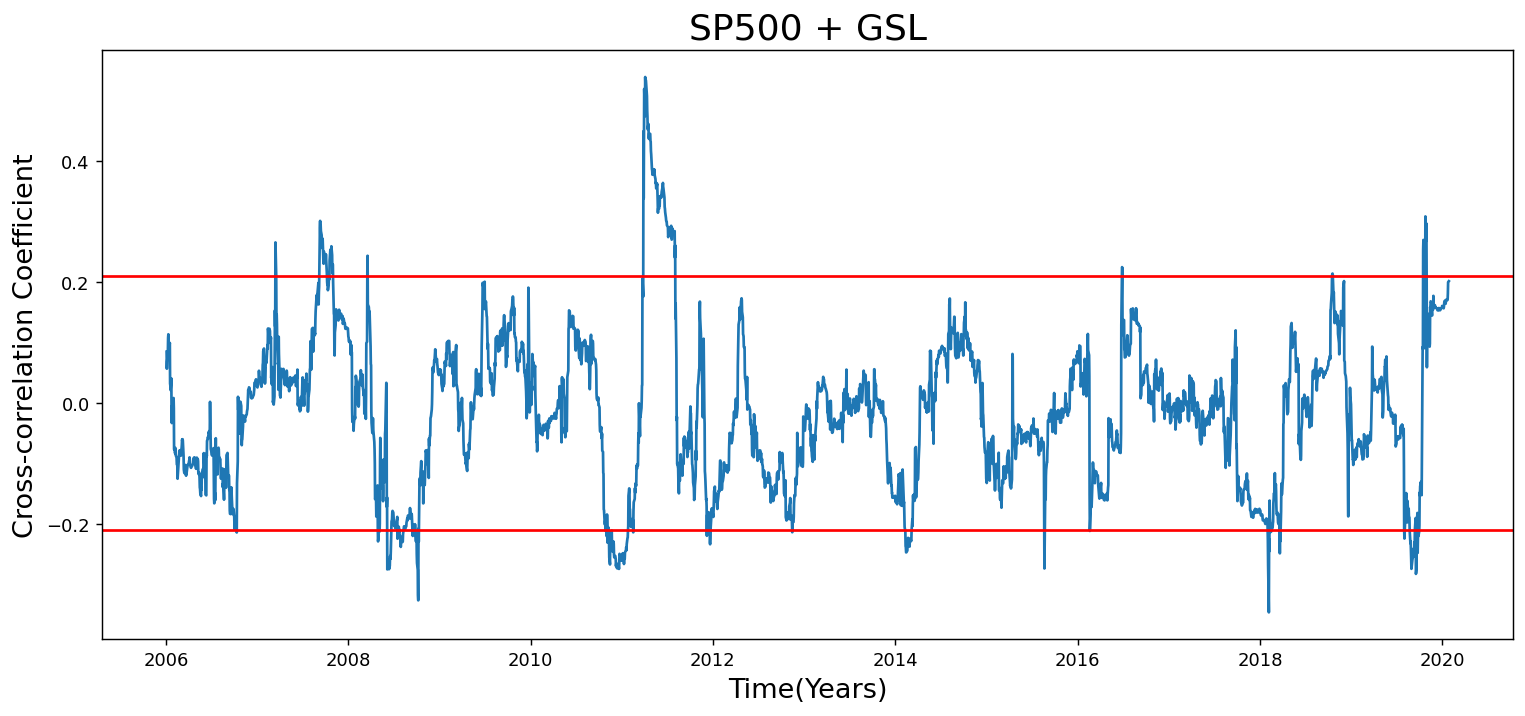}
\includegraphics[scale=.25]{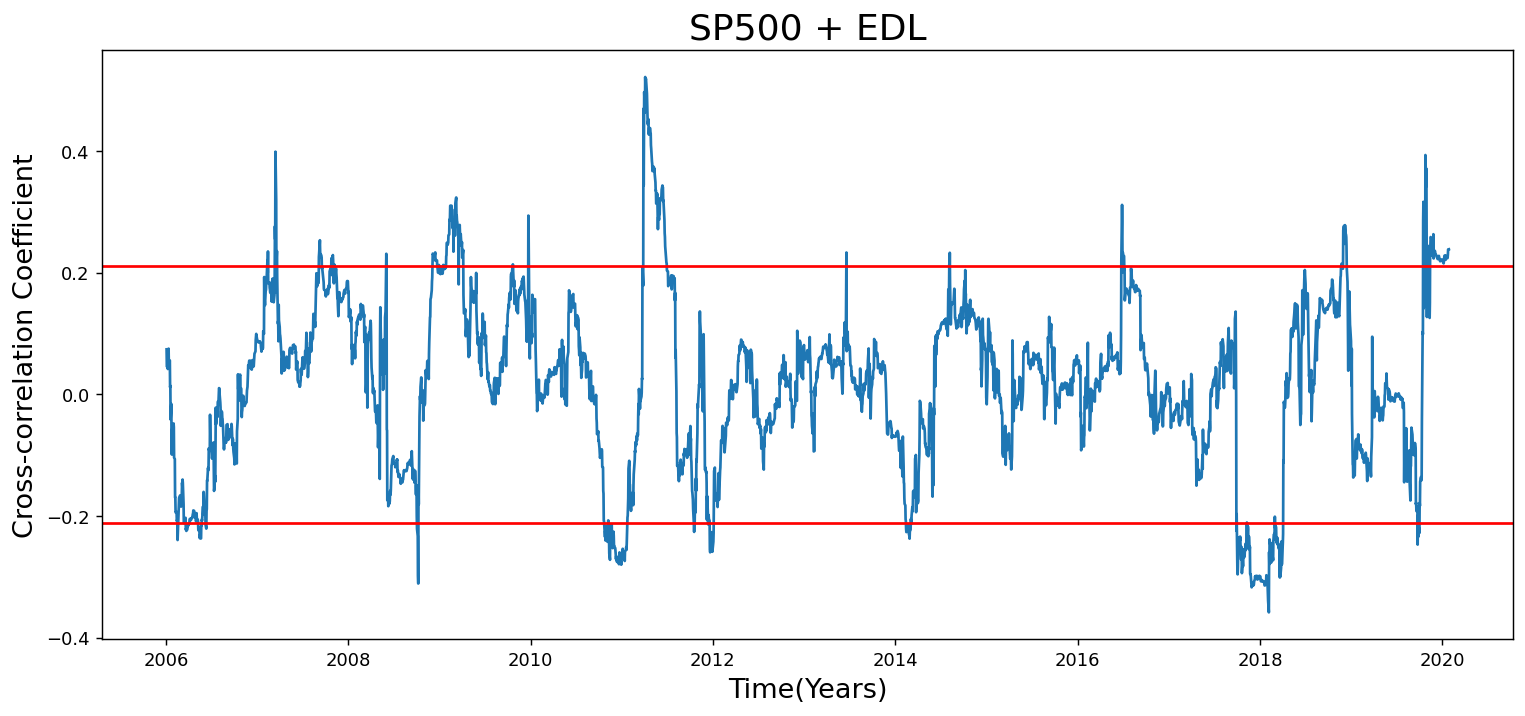}
\includegraphics[scale=.25]{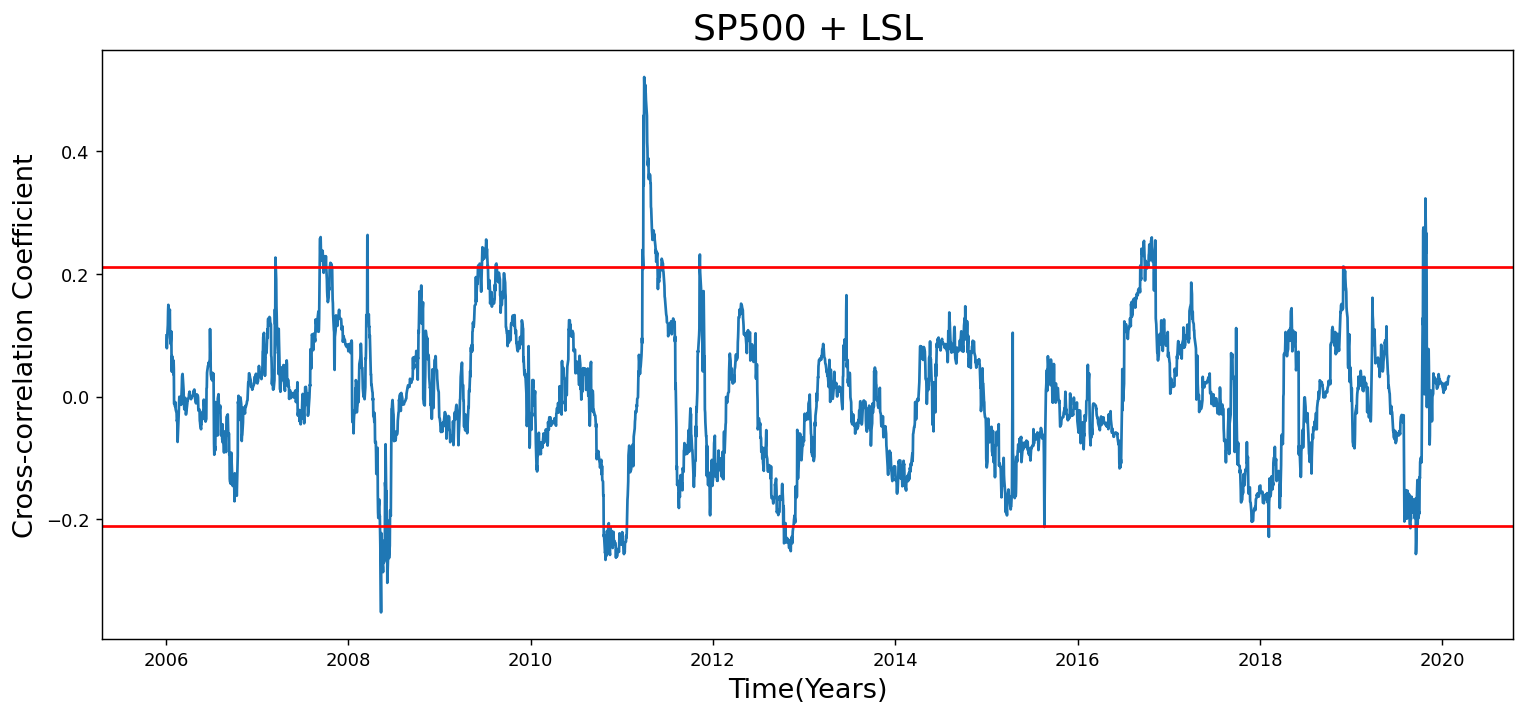}
\includegraphics[scale=.25]{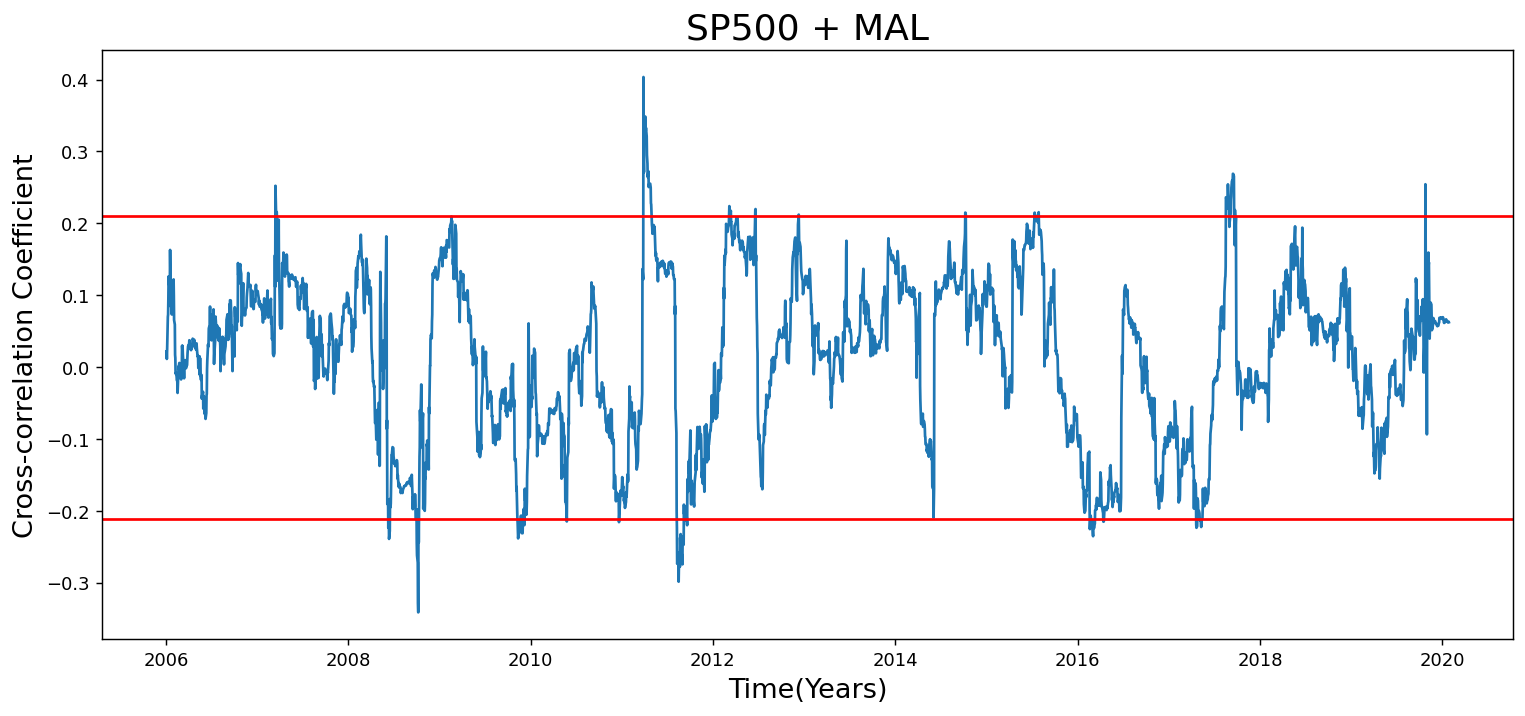}
\includegraphics[scale=.25]{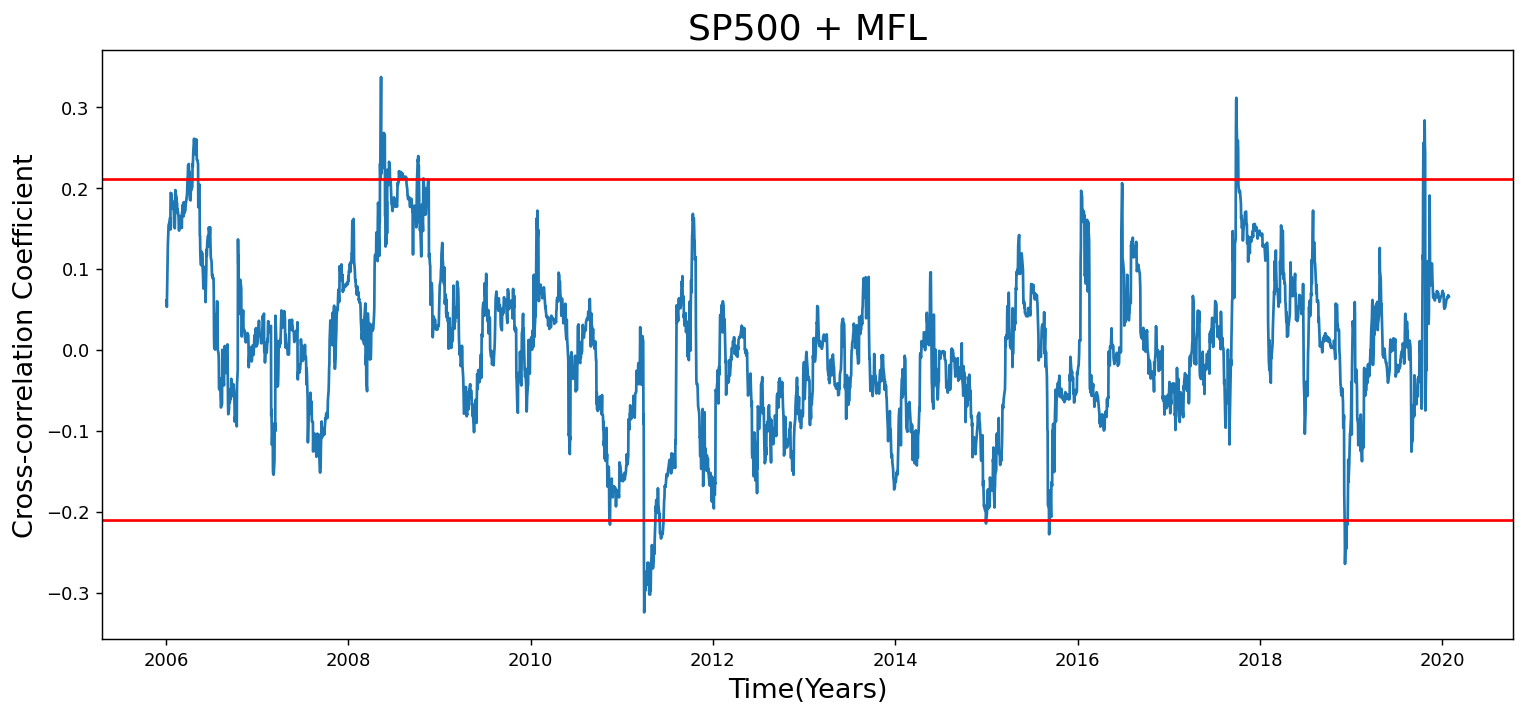}
\caption{Cross-correlation coefficients plot (represented by blue line) with significance bands (red line) between the GSL, EDL, LSL, MAL, and MFL  hedge fund sub-indices and the S\&P 500 over the 174-month time frame from January 2006 to June 2020.}
\label{fig:SP500 & LAB cross-correlations}
\end{figure}

From the 5 plots, it is noticeable that not all sub-indices have similar statistically significant ranges of data similar to the overall LAB index comparison. Therefore, the data represented by the LAB index comparison must mostly be made up of statistically significant data from only the indices that share similar trends. Looking at that statistically significant time period for the LAB index from early 2007 to 2008, that similar frequently positive data outside of the significance bands (~21\% for all comparisons) is present mostly in the GSL (high 20\% to low 30\%), EDL (high 20\% to low 40\%), and a little of the LSL (mid to high 20\%) sub-indices during that same time period. The other 2 indices, the MAL and MFL mostly stay well within the significance bands during that period. Given that the LAB is made up of these sub-indices, we can then see that the statistically high cross-correlation data between the overall hedge fund industry and US market from that 2007-2008 market crash period is mostly created by those 3 hedge fund sectors (GSL, EDL, and LSL). Similar to the analysis of Figure 1, it is also possible that this cross-correlation is an indicator of market friction between those specific hedge fund sectors and the US market (represented by the S\&P 500). Additionally, systemic risk may then also be associated with those sectors. Overall, the results of the cross-correlation calculations between the return data of the LAB and the 5 sub-indices it consists of, with the S\&P 500, seem to show that statistically significant systemic risk was present between the overall hedge fund industry and the US market during the market crash from early 2007-2008. Specifically, the data plots of the LSL, EDL, GSL sub-indices/ hedge fund sectors were mostly responsible for the significant LAB data during that time period, making those sectors the root of that possible systemic risk. 

The GSL index contributes largely to the significant cross-correlations noticed in the LAB, but with this index being a variety of investment types that don't belong in the other categories, it is difficult to pinpoint where/ what investment strategies the possible systemic risk is stemming from and would require a further investigation of the hedge funds included in the sub-indice and their relations with the rest of the economy. Looking at the definition of the LSL index from Section 4.1, it is seen that this group of hedge funds engages in both long and short positions of investing. While long positions are not very risky financial movements, short positions often are as they can be done with extreme leveraging and complex unregulated financial instruments. According to the study conducted by Ferguson and Laster (Date), hedge funds using long/short strategies can maximize possible returns with higher leveraging and the use of obscure new financial instruments. These instruments would often in some way be related to other parts of the economy such as investing in insurance premiums and using money lent from banks. In some way, hedge funds were actually absorbing risk from the financial markets in a possibly healthy way. However, these instruments combined with the lack of regulation by the SEC in the early 2000's allowed hedge funds to take extremely risky and high leverage positions on these investments that involved other institutions as well. Thus, when these hedge funds eventually failed as a result of excessive loss from high risk or sometimes frauds, the various parts of the financial market they were involved with failed as well from the loss. In other words, the LSL strategy hedge funds that had been absorbing a lot of market risk had failed and with that the risk spilled over to other parts of the financial market causing a systemic failure. Similarly, in the Event Driven Liquidity Index (EDL) it can also be seen that hedge funds in these sectors also can develop potentially dangerous relations with other institutions leading to increased systemic risk. As seen in the definition in Section 4.1, the EDL index consists of hedge funds using investment strategies based on the economic movements of short term events such as elections, company mergers, or bankruptcies. In the article written by Teun Johnston (2007) it is shown that these hedge funds have "a dependence on external factors and in the short term, there can be equity market relation", showing that these some of the investment strategies of EDL index members create market friction between themselves and other financial institutions, contributing to systemic risk. Additionally, Johnston also explains that "one of the major risks inherent in event-driven strategies is that managers’ portfolios can be concentrated. Strategies such as merger arbitrage or corporate restructuring are highly correlated to corporate activity and economic cycle", showing how hedge fund involvement in such events creates major interactions between parts of the economy. This possibility of high market friction along with the potentially dangerous concentration in the portfolios of these hedge funds pose a threat of systemic failure as the crashing of these funds can easily spill over to the institutions they are involved with.

\section{Conclusion}

With the advancements our society has made and is making, it is inevitable that our financial systems will become increasingly complex as well. The network between different financial institutions will become more interconnected as technology progresses. These increased connections as well as changes to the regulations of our economy will provide new avenues for systemic risk to develop and pose a threat of systemic failure. Thus, it is crucial that we study this systemic risk and develop a map of its roots and causes to pinpoint the areas of our financial markets that need to have possible systemic prevention measures employed.

The principal component analysis in this paper helps to measure and identify systemic risk through analyzing return data for evidence of market friction across Hedge Funds, Banks, Brokerages, and Insurers. From these tests we were able to conclude that spikes in market friction did occur between these financial institutions suggesting that systemic risk within the entirety of the US financial market was present during this time period and may have been a cause of the total failure. With identifying systemic risk within the US financial markets to exist during crash time, we used cross-correlations to develop a map of possible roots based on hedge funds, given that many other researchers had found hedge funds to be a likely stem of systemic risk. Our first cross-correlation test of the LAB Hedge Fund Index with the S\&P 500 showed statistically significant correlation to occur around crash time, meaning market friction and systemic risk was present between Hedge Funds and the US Economy. We then further defined our map of hedge funds by comparing each of the LAB is 5 sub-indices (EDL, LSL, GSL, MAL, and MFL) to figure out which of those indices were contributing most to the systemic risk found within the entire hedge fund industry. Our results displayed that the GSL, EDL, and LSL sub-indices were the hedge fund sectors mainly accountable for the rise in systemic risk throughout the hedge fund industry which also spread through the entire US financial market.

The results from the tests conducted highlight the need to evaluate the network and connections Hedge Funds have with other hedge funds and the rest of the market, especially the funds part of the GSL, EDL, and LSL indices. Eichengreen and Mathieson (1998) study these connections in relation to the 1997 market crash and Asian currency crash, developing a map of connections between hedge funds and the entire financial market. Their work is relevant in discovering the exact possible areas for the spread of systemic risk, helping to identify where stricter regulation may need to be. On the lines of regulation, having concluded from this study that areas of systemic risk do exist within the hedge funds sector, it is also important to study the regulations in place in these potentially dangerous sectors. Kambhu, Schuermann, and Stiroh (2007) conducted research on this topic amid the market crisis, analyzing the relations between CCRM (Counterparty credit risk management, see section 2) regulations and hedge funds to determine the effectiveness of those measures and to which extent the failures of those measures created the systemic risk and failure. From this study it is observed that changes to CCRM may need to be made in order to prevent hedge funds and other financial institutions from taking potentially dangerous actions such as extreme leveraging and use of complex trading strategies that may not have been addressed.

Our world will always keep advancing, making our financial networks more and more complex. While these improvements are meant for the betterment of all, the complexity of them also creates the possibilities of more threats to our economy. It is crucial that a solid understanding of this increasingly complex network be understood and the necessary regulations be imposed on it to ensure the safety of our economy from complete systemic failures.

\pagebreak

\section{References}

\begin{enumerate}
    \item Adrian, T., \& Brunnermeier, M. K. (2009). "CoVaR". SSRN Electronic Journal. doi:10.2139/ssrn.1269446
    \item Allen, F. (2001). "Do Financial Institutions Matter?" \textit{The Journal of Finance}, 56(4), 1165-1175. doi:10.1111/0022-1082.00361 
    \item Billio, M., Getmansky, M., Lo, A., \& Pelizzon, L. (2010). "Econometric Measures of Systemic Risk in the Finance and Insurance Sectors". doi:10.3386/w16223 
    \item Chan, N., Getmansky, M., Haas, S., \& Lo, A. (2005). "Systemic Risk and 
    Hedge Funds". doi:10.3386/w11200
    \item Chiang, T. C., \& Li, H. (2007). "Dynamic correlation analysis of financial contagion: Evidence from Asian markets". \textit{Journal of International Money and Finance}, 26(7), 1206-1228. doi:10.1016/j.jimonfin.200
    
    7.06.005
    \item De Bandt, Olivier, and Philipp Hartmann, 2000, “Systemic Risk: A Survey,” European Central Bank Working Paper No. 35
    \item Eichengreen, B. J., Mathieson, D. J., \& Chadha, B. (1998). "Hedge funds and financial market dynamics". Washington, DC, District of Columbia: International Monetary Fund.
    \item Ferguson, R., \& Laster, D. (2007). \textit{Hedge Funds and Systemic Risk} (p. 45-51). Banque De France Financial Stability Review - Special Issue on Hedge Funds.
    \item Getmansky, M., Lo, A. W., \& Makarov, I. (2004). "An econometric model of serial correlation and illiquidity in hedge fund returns". \textit{Journal of Financial Economics}, 74(3), 529-609. doi:10.1016/j.jfineco.200
    
    4.04.001
    \item Johnston, T. (2007). Event-Driven Strategies. \textit{The Hedge Fund Journal}, (27).
    \item Kambhu, J., Stiroh, K. J., \& Schuermann, T. (2007). "Hedge Funds, Financial Intermediation, and Systemic Risk". SSRN Electronic Journal. doi:10.2139/ssrn.995907
    \item Kritzman, M., Li, Y., Page, S., \& Rigobon, R. (2011). "Principal Components as a Measure of Systemic Risk". \textit{The Journal of Portfolio Management}, 37(4), 112-126. doi:10.3905/jpm.2011.37.4.112
    \item Lo, A. W. (2008). "Hedge Funds, Systemic Risk, and the Financial Crisis of 2007-2008: Written Testimony for the House Oversight Committee Hearing on Hedge Funds". SSRN Electronic Journal. doi:10.2139/ssrn.1301217
    \item Muirhead, R. (1982). "Aspects of Multivariate Statistical Theory". Wiley Series in Probability and Statistics. doi:10.1002/9780470316559
    \item Rickards, J. (2012, August 7). Repeal of Glass-Steagall caused the Financial Crisis. \textit{U.S. World News and Report}. Retrieved from https://www.usnews.com/opinion/blogs/economic-intelligence/2012/08/27/
    
    repeal-of-glass-steagall-caused-the-financial-crisis
    \item Samuelson, P. A. (2015). "Proof that Properly Anticipated Prices Fluctuate Randomly". The World Scientific Handbook of Futures Markets World Scientific Handbook in Financial Economics Series, 25-38. doi:10.1142/9789814566926\_0002
    \item Singh, M. (2020, July 7). "The 2007-08 Financial Crisis in Review". Retrieved 2020, from https://www.
    
    investopedia.com/articles/economics/09/financial-crisis-review.asp
    
\end{enumerate}

\end{document}